\documentclass[%
 reprint,
superscriptaddress,
 amsmath,amssymb,
 aps,
]{revtex4-2}%{revtex4}

\usepackage{graphicx}% Include figure files
\usepackage{subfigure}
\usepackage{dcolumn}% Align table columns on decimal point
\usepackage{bm}% bold math
\usepackage[colorlinks, linkcolor=blue, urlcolor=blue, anchorcolor=blue, citecolor=blue]{hyperref}
\usepackage{verbatim}
\usepackage{pifont}
\usepackage{float}
\usepackage{enumerate}
\usepackage{multirow}

\begin{document}
\preprint{APS/123-QED}
\title{Practical continuous-variable quantum secret sharing using local local oscillator}
\author{Qin Liao}
\email{llqqlq@hnu.edu.cn}
\affiliation{College of Computer Science and Electronic Engineering, Hunan University, Changsha 410082, China}
\author{Zhuoying Fei}
\affiliation{College of Computer Science and Electronic Engineering, Hunan University, Changsha 410082, China}
\author{Lei Huang}
\affiliation{College of Computer Science and Electronic Engineering, Hunan University, Changsha 410082, China}
\author{Xiquan Fu}
\affiliation{College of Computer Science and Electronic Engineering, Hunan University, Changsha 410082, China}
\date{\today}

\begin{abstract}
Although continuous-variable quantum secret sharing (CVQSS) has been theoretically proven to be secure, it may still be vulnerable to various local oscillator (LO)-aimed attacks. To close this loophole, we propose a practical CVQSS scheme using local LO (LLO), which is called LLO-CVQSS. In this scheme, LO is no longer generated by each user but can be locally generated by the legitimate party, i.e., the dealer. This waives the necessity that all LOs have to be transmitted through an untrusted channel, which makes CVQSS system naturally immune to all LO-aimed attacks, greatly enhancing its practical security. We also develop a specially designed phase compensation method for LLO-CVQSS so that the phase noise of the whole system can be eliminated. We finally construct a noise model for LLO-CVQSS and derive its security bound against both eavesdroppers and dishonest users. Numerical simulation shows that LLO-CVQSS is able to support 30 users at the same time and its maximal transmission distance reaches 112 km, revealing that LLO-CVQSS is not only has the ability to defend itself against all LO-aimed attacks but also has the potential for building large-scale practical quantum communication networks.

%\begin{description}
%\item[Usage]
%\item[Structure]
%\end{description}
\end{abstract}
\maketitle

\section{\label{sec:level1}introduction\protect}
Secret sharing (SS) is originally proposed by Blakely \cite{79Blakley} and Shamir \cite{79Shamir} in 1979, aiming to deliver multiple secret keys to a group of remote users. In a typical ($k$, $n$) threshold SS scheme, the secret is divided into $n$ shares and it can only be fully reconstructed from at least any $k$ of these shares. With the rapid development of quantum technologies, a quantum version of SS (QSS) has been proposed \cite{9903Hillery,9901Karlsson}, whose security is no longer based on computing complexity, but is guaranteed by the laws of quantum mechanics \cite{0907Scarani}, thereby improving the theoretical security level of SS from computational security to unconditional security.  

In general, QSS can be divided into discrete-variable (DV) QSS \cite{9907Cleve,0405LXiao} and continuous-variable (CV) QSS \cite{0503Lance}. The former usually exploits the polarization states of a single photon to transmit the information of key bits \cite{BB84}, while the latter usually encodes key bits in the quadratures ($\hat{x}$ and $\hat{p}$) of the optical field \cite{GG02}. 
In 2002, QSS is extended to CV regime by optical interferometry \cite{0204Tyc}. Subsequently, two more general (2, 3) threshold CVQSS schemes are proposed \cite{0301Lance}, and the efficacy is examined in terms of fidelity, signal transfer coefficients and conditional variances of reconstructed output states. After that, a unified cluster-state QSS for CV with deterministic multi-partite entangled states and high fidelity measurement is put forward \cite{1310Lau}. 
Then, the unconditionally security for CV version of the HBB-type scheme \cite{9903Hillery} against arbitrary attacks is proven \cite{1701Kogias}. However, the above-mentioned CVQSS schemes are all based on entanglement source which is currently difficult to prepare, especially when the number of users $n$ is large. To solve this practical problem, CVQSS using weak coherent states is proposed \cite{1908Grice}. In this scheme, each user injects locally prepared coherent states into a circulating optical mode rather than prepares entanglement states. Subsequently, a CVQSS scheme using discretely modulated coherent states (DMCSs) is further suggested \cite{2103QLiao}, which is easier implementation for CVQSS with existing classical optical modulation technologies. Very recently, CVQSS using multi-ring modulation, which can reduce the error probability of quantum detector for discriminating DMCSs, is proposed, enhancing the performance of DMCS-based CVQSS \cite{2308QLiao}.

Although using coherent states to implement CVQSS is theoretically well studied, another practical issue has been neglected, that is the loophole of transmitting local oscillator (LO). As known, LO is necessary for coherent detection to obtain the measurement results. In the above CVQSS schemes using weak coherent states, LOs are assumed to be generated by each user and have to be transmitted to the dealer. Apparently, a practical CVQSS system may be subject to a series of LO-aimed attacks when LO passes through an insecure channel \cite{2208QLiao}. Till now, researchers have already found several practical attacks against the transmitted LO, such as wavelength attack \cite{1305XCMa,1306JZHuang}, LO calibration attack \cite{1306Jouguet}, LO fluctuation attack \cite{1308XCMa}, clock synchronization jitter \cite{1801CLXie}, and LO polarization attack \cite{1811YJZhao}. These LO-aimed attacks have been investigated in point-to-point continuous-variable quantum key distribution (CVQKD) system \cite{2403YCZhang, 2303ZYChen}, and various countermeasures have been proposed \cite{1403XCMa,1502Kunz,1708WQLiu,1705PHuang,2008YYMao}. However, researches for multi-party CVQSS defending LO-aimed attacks have barely been studied. As a practical CVQSS system is more complicated than a practical CVQKD system due to the increased number of users, it is reasonable to believe that the loophole of transmitting LO in CVQSS system is more severe than that in CVQKD system. 

To solve the above issue, our previous work \cite{2201QLiao} has tentatively suggested a plug-and-play structure for CVQSS where both quantum signal and LO are generated by the receiver. By applying this structure, LO does not needs to be transmitted, thereby avoiding all LO-aimed attacks. However, This advantage comes with the price of transmitting unmodulated quantum signal via an entire untrusted quantum channel, which may introduces new security vulnerabilities. We therefore are still seeking a better method for closing the loophole of transmitted LO in CVQSS without introducing other troubles. 

Fortunately, there exists a local LO (LLO) countermeasure for point-to-point CVQKD which is immune to all LO-aimed attacks \cite{1510BQi}. Over the past few years, CVQKD using LLO has been theoretically and experimentally investigated \cite{1508DHuang,1802TWang,2010HWang}, and the results have shown its feasibility in practical CVQKD system. Inspired by these solutions, in this paper, we extend LLO technique from point-to-point quantum communication scenario to multi-party quantum communication scenario and thus propose a practical CVQSS scheme based on LLO (LLO-CVQSS), aiming to thoroughly close the loophole that transmitted LOs of CVQSS system may be subject to various attacks. In particular, we first present the detailed design of LLO-CVQSS in which all LOs are no longer required to be transmitted via untrusted channel, so that the practical security of CVQSS system can be improved. Then a specially designed phase compensation method for multi-party quantum communication system is also developed so that phase noise of LLO-CVQSS system caused by both fast and slow phase drifts can be eliminated. We finally construct a noise model for the proposed LLO-CVQSS system and derive its security bound against both eavesdroppers and dishonest users. Numerical simulation shows that the maximal transmission distance of LLO-CVQSS can be reached to 112 km and it can support up to 30 users to sharing secret keys at the same time, revealing that the proposed LLO-CVQSS is not only has the ability to defend itself against all LO-aimed attacks but also has the potential for building large-scale practical quantum communication networks.

This paper is organized as follows. In Sec. \ref{II}, we detail the design of LLO-CVQSS, phase compensation method and noise model. Performance analysis and discussion are presented in Sec. \ref{III}. Conclusions are drawn in Sec. \ref{IV}.

\section{\label{II}LLO-CVQSS scheme\protect}
Here we present the main steps of the proposed LLO-CVQSS scheme, its phase compensation method and noise model are subsequently detailed. For simplicity, we only focus on the basic (2, 2) threshold LLO-CVQSS scheme in which two remote users respectively share a part of secret key with the dealer. The more complicated ($n$, $n$) threshold LLO-CVQSS can be further derived with similar ideas. 
\subsection{\label{IIA}Design of LLO-CVQSS}
\begin{figure*}[!htbp]
\centering
\includegraphics[width=17.5cm]{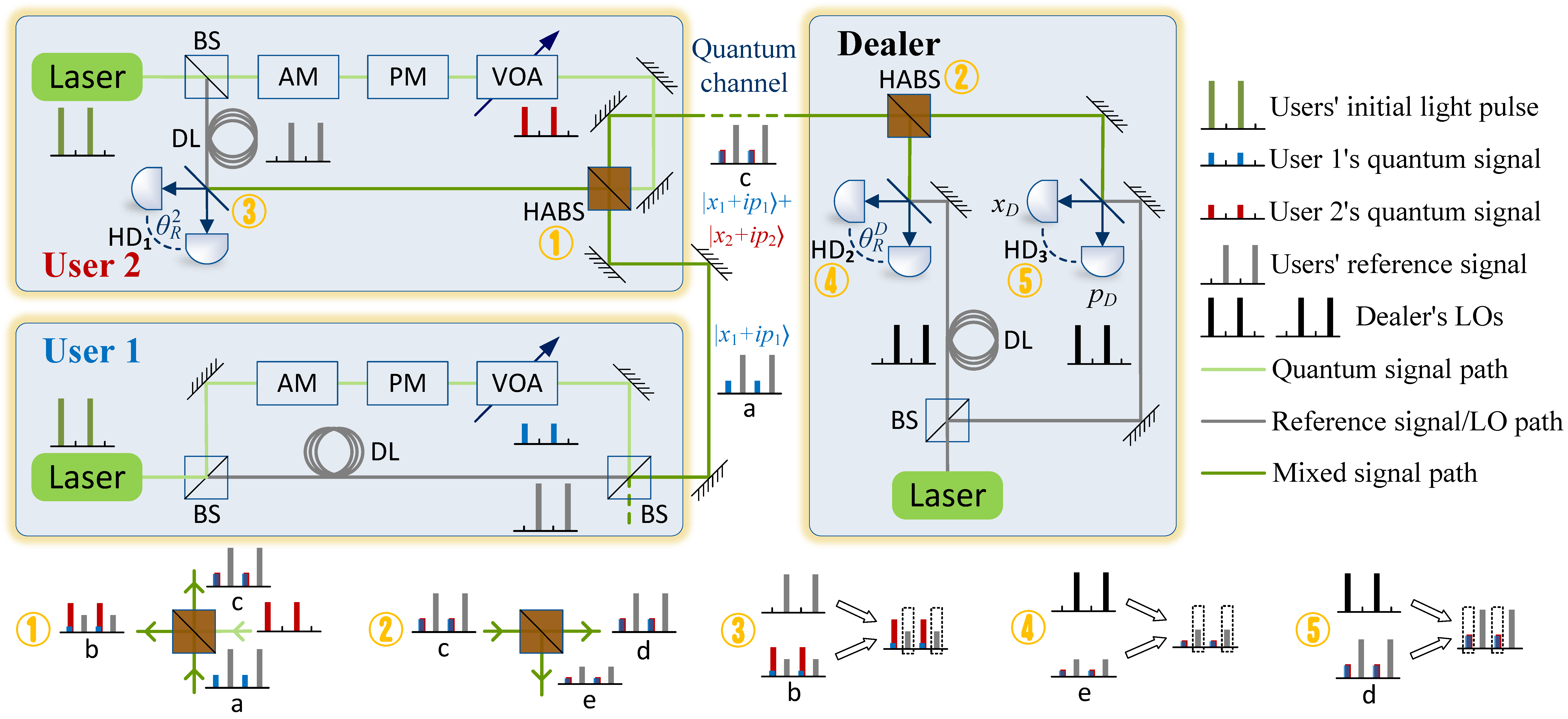}
\caption{Schematic diagram of (2, 2) threshold LLO-CVQSS in which two remote users (user 1 and user 2) are orderly connected with the dealer through an unstrusted quantum channel. Each user locally generates GMCSs and sends them to the dealer who performs heterodyne detection with his own locally generated LO. The specific operations of \ding{172}-\ding{176} are indicated below and the pulse legends are on the right. BS, Beam splitter; DL, Delay line; AM, Amplitude modulator; PM, Phase modulator; VOA, Variable optical attenuator; HABS, Highly asymmetric beam splitter; HD, Heterodyne detector.}
\label{fig:Program}
\end{figure*}
As shown in Fig. \ref{fig:Program}, two remote users (user 1 and user 2) are orderly connected with the dealer through an unstrusted quantum channel, and each user locally generates the Gaussian-modulated coherent states (GMCSs) and sends them to the dealer. Its main steps are described below.

\emph{Step 1}. User 1 generates light pulse with his own laser and subsequently splits it into two paths through a 50:50 beam splitter (BS). The GMCS is prepared by one of the paths after passing amplitude modulator (AM), phase modulator (PM) and variable optical attenuator (VOA), while user 1's reference signal is formed by another path after delaying half an interval. These two kinds of signals are merged by a BS and then sent to user 2, the output quantum state of user 1 can be represented as $|x_1+ip_1\rangle$.

\emph{Step 2}. User 2 also generates light pulse with his own laser and splits it into two paths through a BS. The GMCS is prepared by one of the paths, while user 2's reference signal is formed by another path after delaying half an interval. The GMCS is coupled to the same spatiotemporal mode as user 1's incoming signal via a highly asymmetric beam splitter (HABS) \cite{1908Grice}. This mixed signal (including $|x_1+ip_1\rangle$, $|x_2+ip_2\rangle$ and user 1's reference signal) will subsequently be sent to the dealer. Besides, user 2's reference signal and the reflected user 1's incoming signal are measured by a heterodyne detector (HD) to obtain the reference phase $\theta_2^R$ between the lasers of user 1 and user 2.

\emph{Step 3}. The dealer locally generates LO and then splits it into two paths by a BS, one of the paths is subsequently delayed with half an interval. The incoming mixed signal is also splitted into two parts by a HABS, one part is measured with the delayed LO to obtain the reference phase $\theta_D^R$ between the lasers of user 1 and the dealer, and the another part is detected with the undelayed LO for obtaining heterodyne measurement results \{$x_D, p_D$\}.

\emph{Step 4}. User 2 and the dealer individually compensate phase fast drifts according to their respective reference phases $\theta_2^R$ and $\theta_D^R$, so that the data of user 2 and the dealer can be respectively revised to \{$x_2', p_2'$\} and \{$x_D', p_D'$\}.

\emph{Step 5}. The above process is repeated many times to acquire sufficient data. 
User 1 and user 2 individually compensate their respective phase slow drifts so that the data of user 1 and user 2 can be respectively revised to $\{x_1', p_1'\}_{m}$, $\{x_2^{''}$, $p_2^{''}\}_{m}$. As a result, user 1, user 2 and the dealer respectively hold related raw data $\{x_1', p_1'\}_{m}$, $\{x_2^{''}, p_2^{''}\}_{m}$ and $\{x_D', p_D'\}_{m}$.

\emph{Step 6}. The dealer randomly selects a subset of his raw data and requests all users to disclose their corresponding raw data, so that the channel transmittance of each user \{$T_{1}, T_{2}$\} can be estimated. Note that this disclosed data have to be discarded after this step.

\emph{Step 7}. Assuming that user 1 is honest and user 2 is dishonest. The dealer further selects another subset of his raw data and requests user 2 to announce the corresponding values, so that the dealer can replace his subset to $x_{D_1}=x_D'-\sqrt{T_2}x_2^{''}$ and $p_{D_1}=p_D'-\sqrt{T_2}p_2^{''}$. In this way, a point-to-point CVQKD link between the dealer and user 1 is actually established. Therefore, the lower bound of the secret key rate $R_1$ between the dealer and user 1 can be estimated by taking advantages of the CVQKD's security proof \cite{0611Nava,0611Gar}. After that, all participants discard the disclosed data. Similarly, the lower bound of the secret key rate $R_2$ between the dealer and user 2 can also be estimated. For security reasons, the dealer should choose the smallest value among \{$R_1, R_2$\} as the final secret key rate $R$ of LLO-CVQSS scheme.

\emph{Step 8}. If the value of $R$ is positive, the secret key $K_1$ between user 1 and the dealer, as well as the secret key $K_2$ between user 2 and the dealer, can be respectively generated by applying standard postprocessing procedures of CVQKD \cite{2402TWang} to the remaining undisclosed data.

\emph{Step 9}. Finally, the dealer encodes the message $M$ according to the expression $E=M \oplus K_1 \oplus K_2$ and then broadcasts the encrypted message $E$ to all users. Obviously, only if user 1 and user 2 work cooperatively can the encrypted message $E$ be decoded.  

Compared with existing CVQSS schemes, our proposed LLO-CVQSS scheme is naturally immune to all LO-aimed attacks. This is because LOs are no longer need to be transmitted to the untrusted channel by every user, but directly generated and measured at the dealer's side. This strategy of locally generating LO from a legitimate measurer eliminates the possibility of LO being attacked, thereby enhancing the practical security of CVQSS protocol. It is worth noting that phase compensations mentioned in \emph{Step 4} and \emph{Step 5} are crucial for implementing our scheme. In what follows, we detail the phase compensation method for the proposed LLO-CVQSS system against phase drift.

\subsection{\label{IIB}Phase compensation method}

\begin{figure*}[!htbp]
\centering
\subfigure{\label{fig:Phase1}
\includegraphics[width=5.7cm]{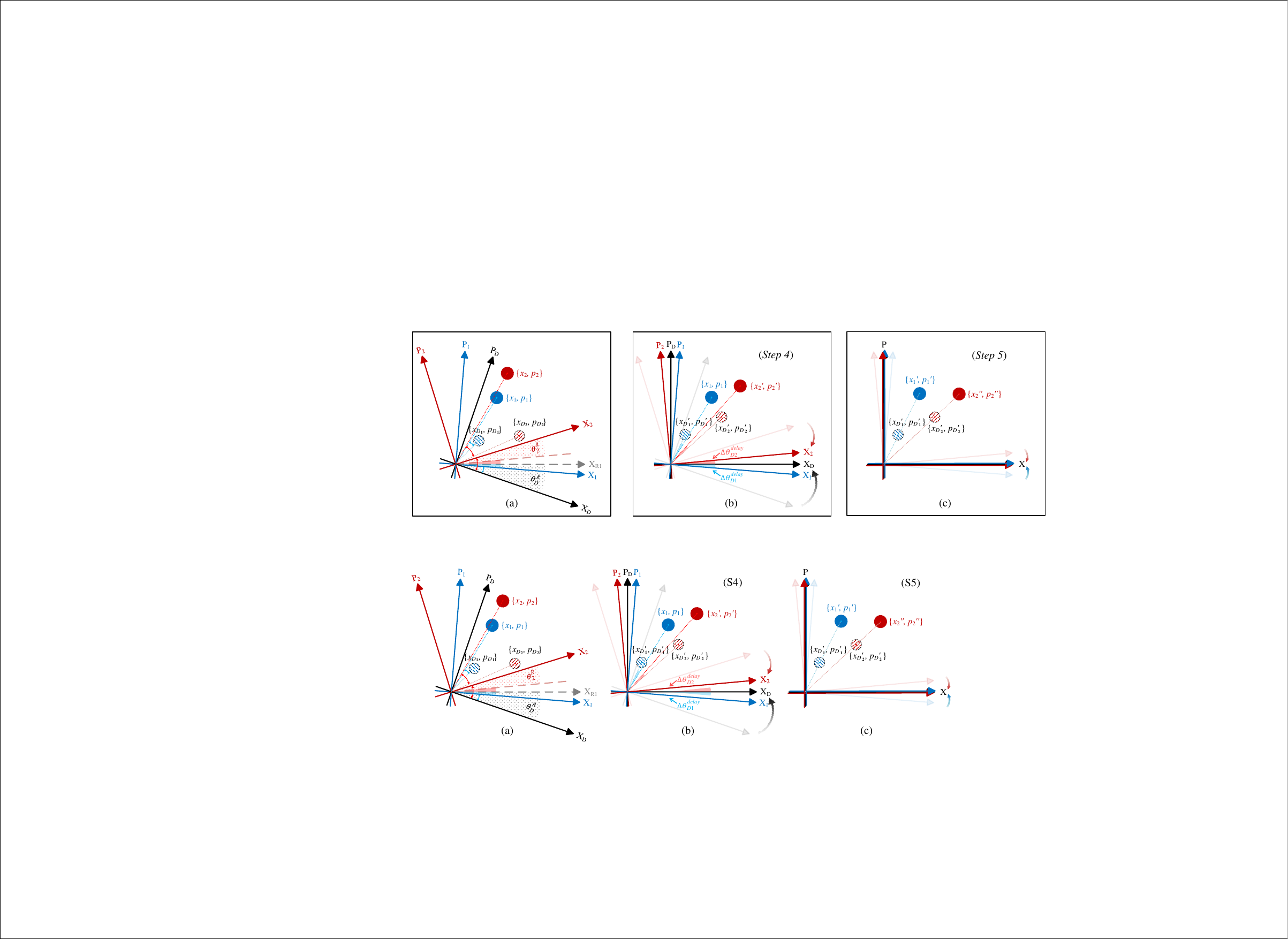}}
\subfigure{\label{fig:Phase2}
\includegraphics[width=5.7cm]{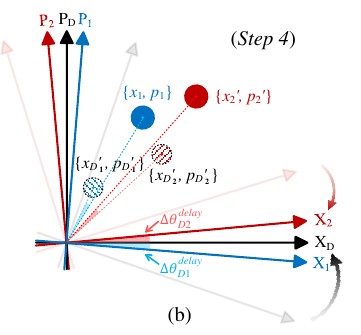}}
\subfigure{\label{fig:Phase3}
\includegraphics[width=5.7cm]{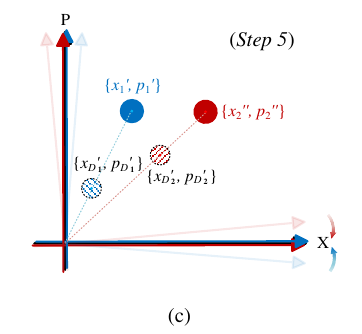}}
\caption{An example for phase compensation method (from dealer's side). The blue, red and black coordinate axes respectively denote the phase bases of user 1, user 2 and the dealer. The blue and red dots are the quantum signals of user 1 and user 2 respectively, and the gray dashed arrow is user 1's reference signal phase. The dashed circles indicate the two parts of dealer's measurement \{$x_{D_1}+x_{D_2}, p_{D_1}+p_{D_2}$\}. (a) LLO-CVQSS without any phase compensation. (b) LLO-CVQSS after compensating phase fast drift. (c) LLO-CVQSS after compensating phase fast and slow drifts.}
\label{fig:Reference system}
\end{figure*}

In general, the phase drift of a practical optical communication system is mainly caused by two aspects, the one is phase fast drift which is aroused by the initial phase difference between independent lasers, and the another one is phase slow drift which is aroused by the transmission delay of optical path and the clock synchronization error of the laser. These two kinds of phase drifts will jointly result in measurement error, severely degenerating the performance of CVQSS system. 
Fig. \ref{fig:Reference system} shows an example for our phase compensation method. For comparison, LLO-CVQSS without applying any phase compensation is first depicted in Fig. \ref{fig:Phase1}. It can be seen that there exists large different phase differences between users' quantum signals and dealer's measurements due to the phase drift. In this situation, dealer's measurement \{$x_D, p_D$\} can be expressed as
\begin{equation}\label{e1}
\begin{aligned}
\Big[\begin{array}{cc}
    x_D\\
    p_D
\end{array}\Big]
=&\Big[\begin{array}{cc}
    x_{D_1}+x_{D_2}\\
    p_{D_1}+p_{D_2}
\end{array}\Big]\\
=&\sqrt{\frac{\eta T_1}2}
\Big[\begin{array}{cc}
    \cos\Phi_1 & \sin\Phi_1 \\
    -\sin\Phi_1 & \cos\Phi_1
\end{array}\Big]
\Big[\begin{array}{cc}
    x_1\\
    p_1
\end{array}\Big]+\\
&\sqrt{\frac{\eta T_2}2}
\Big[\begin{array}{cc}
    \cos\Phi_2 & \sin\Phi_2 \\
    -\sin\Phi_2 & \cos\Phi_2
\end{array}\Big]
\Big[\begin{array}{cc}
    x_2\\
    p_2
\end{array}\Big]
+\Big[\begin{array}{cc}
    x_N\\
    p_N
\end{array}\Big],
\end{aligned}
\end{equation}
where $\eta$ is the efficiency of heterodyne detector, $x_N$ and $p_N$ are the excess noise, $\Phi_{1}$ and $\Phi_{2}$ are the phase drifts of user 1 and user 2's quantum signals, which can be respectively expressed as
\begin{equation}\label{e2}
\Phi_1=\theta_1^{init}-\theta_D^{init}
\end{equation}
and
\begin{equation}\label{e3}
\Phi_2=\theta_2^{init}-\theta_D^{init},
\end{equation}
where $\theta_{1}^{init}$, $\theta_{2}^{init}$ and $\theta_{D}^{init}$ are the initial phases of user 1, user 2 and dealer's laser. By introducing these initial phases, the reference phases $\theta_2^R$ and $\theta_D^R$, which are respectively measured by user 2 (\emph{Step 2}) and the dealer (\emph{Step 3}), can be derived by
\begin{equation}\label{e4}
\theta_2^R=\theta_1^{init}+\theta_1^{delay}-(\theta_2^{init}+\theta_2^{delay})
\end{equation}
and
\begin{equation}\label{e5}
\theta_D^R=\theta_1^{init}+\theta_1^{delay}-(\theta_D^{init}+\theta_D^{delay}),
\end{equation}
where $\theta_{1}^{delay}$ is cumulative phase of user 1's reference signal, $\theta_{2}^{delay}$ is cumulative phase of user 2's reference signal, and $\theta_{D}^{delay}$ is cumulative phase of dealer's delayed LO.

Combining Eq.(\ref{e4}) and Eq.(\ref{e5}), Eq.(\ref{e2}) and Eq.(\ref{e3}) can be finally rewritten to 
\begin{equation}\label{e6}
\begin{aligned}
\Phi_1&=\theta_D^R-\theta_1^{delay}+\theta_D^{delay}\\
&=\theta_D^R+\Delta\theta_{D1}^{delay}
\end{aligned}
\end{equation}
and
\begin{equation}\label{e7}
\begin{aligned}
\Phi_2&=\theta_D^R-\theta_2^R-\theta_2^{delay}+\theta_D^{delay}\\
&=\theta_D^R-\theta_2^R+\Delta\theta_{D2}^{delay},
\end{aligned}
\end{equation}
where $\Delta\theta_{D1}^{delay}$ is the cumulative phase difference between the dealer and user 1, $\Delta\theta_{D2}^{delay}$ is the cumulative phase difference between the dealer and user 2. From Eq.(\ref{e6}) and Eq.(\ref{e7}), we can explicitly tell that the total phase drift of each user's quantum signal consists of two components, namely the user's reference phase difference with the dealer, and the user's cumulative phase difference with the dealer. Therefore, we can perform two phase rotations to respectively compensate these two kinds of phase differences.

For the first phase rotation, both user 2 and the dealer locally revise their data according to their respective obtained reference phases $\theta_2^R$ and $\theta_D^R$. Specifically, user 2 modifies his data \{$x_2, p_2$\} to \{$x_2', p_2'$\} with the equation
\begin{equation}\label{e8}
\begin{aligned}
\Big[\begin{array}{cc}
    x_2'\\
    p_2'
\end{array}\Big]
=\Big[\begin{array}{cc}
    \cos(-\theta_2^R) & \sin(-\theta_2^R) \\
    -\sin(-\theta_2^R) & \cos(-\theta_2^R)
\end{array}\Big]
\Big[\begin{array}{cc}
    x_2\\
    p_2
\end{array}\Big],
\end{aligned}
\end{equation}
and the dealer modifies his measurement result \{$x_D, p_D$\} to \{$x_D', p_D'$\} with the equation
\begin{widetext}\begin{equation}\label{e9}
\begin{aligned}
\Big[\begin{array}{cc}
    x_D'\\
    p_D'
\end{array}\Big]
=&\Big[\begin{array}{cc}
    \cos(-\theta_D^R) & \sin(-\theta_D^R) \\
    -\sin(-\theta_D^R) & \cos(-\theta_D^R)
\end{array}\Big]
\Big[\begin{array}{cc}
    x_D\\
    p_D
\end{array}\Big]\\
=&\sqrt{\frac{\eta T_1}2}
\Big[\begin{array}{cc}
    \cos(\Phi_1-\theta_D^R) & \sin(\Phi_1-\theta_D^R) \\
    -\sin(\Phi_1-\theta_D^R) & \cos(\Phi_1-\theta_D^R)
\end{array}\Big]
\Big[\begin{array}{cc}
    x_1\\
    p_1
\end{array}\Big]+
\sqrt{\frac{\eta T_2}2}
\Big[\begin{array}{cc}
    \cos(\Phi_2-\theta_D^R) & \sin(\Phi_2-\theta_D^R) \\
    -\sin(\Phi_2-\theta_D^R) & \cos(\Phi_2-\theta_D^R)
\end{array}\Big]
\Big[\begin{array}{cc}
    x_2\\
    p_2
\end{array}\Big]
+\Big[\begin{array}{cc}
    x_N'\\
    p_N'
\end{array}\Big]\\
=&\sqrt{\frac{\eta T_1}2}
\Big[\begin{array}{cc}
    \cos(\Delta\theta_{D1}^{delay}) & \sin(\Delta\theta_{D1}^{delay}) \\
    -\sin(\Delta\theta_{D1}^{delay}) & \cos(\Delta\theta_{D1}^{delay})
\end{array}\Big]
\Big[\begin{array}{cc}
    x_1\\
    p_1
\end{array}\Big]+
\sqrt{\frac{\eta T_2}2}
\Big[\begin{array}{cc}
    \cos(-\theta_2^R+\Delta\theta_{D2}^{delay}) & \sin(-\theta_2^R+\Delta\theta_{D2}^{delay}) \\
    -\sin(-\theta_2^R+\Delta\theta_{D2}^{delay}) & \cos(-\theta_2^R+\Delta\theta_{D2}^{delay})
\end{array}\Big]
\Big[\begin{array}{cc}
    x_2\\
    p_2
\end{array}\Big]\\
&+\Big[\begin{array}{cc}
    x_N'\\
    p_N'
\end{array}\Big],
\end{aligned}
\end{equation}
\end{widetext}
where $x_N'$ and $p_N'$ are Gaussian noise. By doing this rotation, the initial phases of both user 2 and the dealer's lasers can be deemed to be identical with that of user 1's laser, which is illustrated in Fig. \ref{fig:Phase2}, thereby eliminating phase fast drift caused by different lasers. However, it can be seen that there still exists cumulative phase differences $\Delta\theta_{D1}^{delay}$ and $\Delta\theta_{D2}^{delay}$ between the dealer and each user, which will be compensated by the second phase rotation.

Different from phase fast drift compensation, cumulative phase differences $\Delta\theta_{D1}^{delay}$ and $\Delta\theta_{D2}^{delay}$ cannot be directly obtained by measurement, we therefore have to seek for another way to compensate these two phase differences. Considering that the slow fluctuating characteristics of cumulative phase difference \cite{1601DHuang}, former data can be used for estimating the cumulative phase difference of latter data. To do so, a set of data needs to be first obtained by repeating previous steps several times. The dealer subsequently discloses a part of his data, so that each user can locally calculate the correlation \cite{1800YJZhao} between himself and the dealer. Specifically, the correlation between user 1 and the dealer can be calculated by
\begin{equation}\label{e10}
\langle x_1x_D'\rangle=\sqrt{\frac{\eta T_1}2}\cos(\Delta\theta_{D1}^{delay})\langle x_1x_1\rangle
\end{equation}
and
\begin{equation}\label{e11}
\langle p_1x_D'\rangle=\sqrt{\frac{\eta T_1}2}\sin(\Delta\theta_{D1}^{delay})\langle x_1x_1\rangle,
\end{equation}
while the correlation between user 2 and the dealer can be calculated by
\begin{equation}\label{e12}
\langle x_2'x_D'\rangle=\sqrt{\frac{\eta T_2}2}\cos(\Delta\theta_{D2}^{delay})\langle x_2x_2\rangle
\end{equation}
and
\begin{equation}\label{e13}
\langle p_2'x_D'\rangle=\sqrt{\frac{\eta T_2}2}\sin(\Delta\theta_{D2}^{delay})\langle x_2x_2\rangle.
\end{equation}
See Appendix \ref{Appendix A} for the detailed derivation. With above equations, user 1 calculates the estimate of $\Delta\theta_{D1}^{delay}$ by
\begin{equation}\label{e14}
\Delta\hat{\theta}_{D1}^{delay}=\arctan\frac{\langle p_1x_D'\rangle}{\langle x_1x_D'\rangle}.
\end{equation}
Likewise, user 2 also calculates the estimate of $\Delta\theta_{D2}^{delay}$ by
\begin{equation}\label{e15}
\Delta\hat{\theta}_{D2}^{delay}=\arctan\frac{\langle p_2'x_D'\rangle}{\langle x_2'x_D'\rangle}.
\end{equation}
Subsequently, user 1 modifies his remaining data $\{x_1, p_1\}$ to $\{x_1', p_1'\}$ with the equation 
\begin{equation}\label{e16}
\begin{aligned}
\Big[\begin{array}{cc}
    x_1'\\
    p_1'
\end{array}\Big]
=\Big[\begin{array}{cc}
    \cos(\Delta\hat{\theta}_{D1}^{delay}) & \sin(\Delta\hat{\theta}_{D1}^{delay}) \\
    -\sin(\Delta\hat{\theta}_{D1}^{delay}) & \cos(\Delta\hat{\theta}_{D1}^{delay})
\end{array}\Big]
\Big[\begin{array}{cc}
    x_1\\
    p_1
\end{array}\Big],
\end{aligned}
\end{equation}
and user 2 modifies his remaining data $\{x_2', p_2'\}$ to $\{x_2'', p_2''\}$ with the equation 
\begin{equation}\label{e17}
\begin{aligned}
\Big[\begin{array}{cc}
    x_2^{''}\\
    p_2^{''}
\end{array}\Big]
=&\Big[\begin{array}{cc}
    \cos(\Delta\hat{\theta}_{D2}^{delay}) & \sin(\Delta\hat{\theta}_{D2}^{delay}) \\
    -\sin(\Delta\hat{\theta}_{D2}^{delay}) & \cos(\Delta\hat{\theta}_{D2}^{delay})
\end{array}\Big]
\Big[\begin{array}{cc}
    x_2'\\
    p_2'
\end{array}\Big]\\
=&\Big[\begin{array}{cc}
    \cos(-\theta_2^R+\Delta\hat{\theta}_{D2}^{delay}) & \sin(-\theta_2^R+\Delta\hat{\theta}_{D2}^{delay}) \\
    -\sin(-\theta_2^R+\Delta\hat{\theta}_{D2}^{delay}) & \cos(-\theta_2^R+\Delta\hat{\theta}_{D2}^{delay})
\end{array}\Big]\\ 
&\times\Big[\begin{array}{cc}
    x_2\\
    p_2
\end{array}\Big].
\end{aligned}
\end{equation}
Eventually, each user's total phase drift can be compensated after performing the second phase rotation, as shown in Fig. \ref{fig:Phase3}.

Note that there is a difference between the above two phase rotations, that is, user 1's data remains unchanged during the first phase rotation, while the dealer's data remains unchanged during the second phase rotation. These different rotating strategies allow each participant locally revise his own data without revealing extra useful information, thereby further ensuring the practical security of LLO-CVQSS.
\subsection{\label{IIC}Noise model for LLO-CVQSS}
Since our proposed scheme is very different from the conventional CVQSS in terms of system architecture and processing steps, a noise model for LLO-CVQSS is needed to be constructed as follows.

In \emph{Step 1} and \emph{Step 2}, both user 1 and user 2 are required to prepare respective GMCS, the modulation variance can not be precisely controlled due to the finite dynamics of practical AMs \cite{2109YShao}. The introduced modulation noise $\varepsilon_{AM}$ can be quantized as
\begin{equation}\label{e18}
\begin{aligned}
\varepsilon_{AM}=&\varepsilon_{AM_1}+\varepsilon_{AM_2}\\
=&|\alpha_{Smax_1}|^2 10^\frac{-d_{dB}}{10}+\frac{T_2}{T_1}\cdot|\alpha_{Smax_2}|^2 10^\frac{-d_{dB}}{10},
\end{aligned}
\end{equation}
where $d_{dB}$ is the finite dynamic of the AM, $\alpha_{Smax_1}$ and $\alpha_{Smax_2}$ are the maximum amplitudes of user 1 and user 2's signals \cite{2304Shen}, which are related to their respective modulation variances $V_{U_1}$ and $V_{U_2}$. The channel transmittance of each user can be presented as $T=10^\frac{-\alpha l}{10}$, where $\alpha$ is the fiber attenuation coefficient and $l$ is the fiber length to the dealer.

Before sending signals to the quantum channel, both user 1 and user 2's modulated optical pulses have to be coupled to user 1's reference signal so that the quantum signals may be contaminated by the residual photons of user 1's reference signal, thereby introducing photon-leakage noise $\varepsilon_{LE}$, which can be quantized as \cite{2203YShao}
\begin{equation}\label{e19}
\begin{aligned}
\varepsilon_{LE}=&\varepsilon_{LE_1}+\varepsilon_{LE_2}\\
=&\frac{2|\alpha_{R_1}|^2}{R_e}+\frac{T_2}{T_1}\cdot\frac{2|\alpha_{R_1}|^2}{R_e},
\end{aligned}
\end{equation}
where $|\alpha_{R_1}|^2$ is the amplitude of user 1's reference signal, $R_e$ is the finite extinction ratio of the AM.

In \emph{Step 3}, the output voltage at dealer’s side is quantized by the imperfect analog-to-digital converter (ADC), which introduces the ADC quantization noise in \{$x_D, p_D$\} \cite{2108Kish}, noted as
\begin{equation}\label{e20}
\begin{aligned}
\varepsilon_{ADC}=&\varepsilon_{ADC_1}+\varepsilon_{ADC_2}\\
\geq&\frac{|\alpha_{Smax_1}|^2}{12\cdot2^q}+\frac{T_2}{T_1}\cdot\frac{|\alpha_{Smax_2}|^2}{12\cdot2^q},
\end{aligned}
\end{equation}
where $q$ is the quantization bit number of the ADC.

Another non-negligible noise is phase noise which can be identified as \cite{2105SJRen,2206HWang}
\begin{equation}\label{e21}
\varepsilon_{phase}=\varepsilon_{fast}+\varepsilon_{slow},
\end{equation}
where compensation noise of phase fast drift $\varepsilon_{fast}=\varepsilon_{drift}+\varepsilon_{error}$ and compensation noise of phase slow drift $\varepsilon_{slow}$ depends on the compensation error $V_{slow}$ \cite{1701Marie}. In the proposed LLO-CVQSS, $\varepsilon_{drift}$ equals 0 since both quantum signal and reference signal of each user are generated by the same laser \cite{2103XDWu}.
$\varepsilon_{error}$ reflects the measurement error of reference phase introduced by user 2 and dealer's heterodyne detection \cite{2207YShao}, which can be calculated as
\begin{equation}\label{e22}
\begin{aligned}
\varepsilon_{error}=&\varepsilon_{error_D}+\varepsilon_{error_2}\\
=&V_{A_1}\frac{\chi_D+1}{|\alpha_{R_1}|^2}+\frac{T_2}{T_1}\cdot V_{A_2}\frac{\chi_2+1}{|\alpha_{R_1}|^2},
\end{aligned}
\end{equation}
where $\chi_D=\frac{2-\eta T_1+2v_{el}}{\eta T_1}+\varepsilon_{ch}$ and $\chi_2=\frac{2-\eta T_2+2v_{el}}{\eta T_2}+\varepsilon_{ch}$ are the total noise imposed on user 1's reference signal, in which $\varepsilon_{ch}$ is channel noise of phase reference \cite{1304Jouguet} and $v_{el}$ is electrical noise.

Up to now, the total excess noise referred to the channel input can be modeled by the above elements as
\begin{equation}\label{e23}
\varepsilon=\varepsilon_{AM}+\varepsilon_{LE}+\varepsilon_{ADC}+\varepsilon_{phase}+\varepsilon_{rest},
\end{equation}
where $\varepsilon_{rest}$ is the other noise that weakens system performance in conventional CVQKD \cite{1801TWang}.
\section{\label{III}Performance analysis and discussion\protect}
For now, the proposed LLO-CVQSS scheme including phase compensation method and its specific noise model have been detailedly presented. Its biggest merit is of course the enhancement of practical security of CVQSS system as LOs can be locally generated without transmitting through untrusted channel. In what follows, we quantificationally evaluate the performance of our proposed LLO-CVQSS scheme.
\begin{table}[h!]
\centering
\renewcommand{\arraystretch}{1.5}
\newcommand{\tabincell}[2]{\begin{tabular}{@{}#1@{}}#2\end{tabular}}
\caption{Global parameters for numerical simulations.}\label{TABLE}
\begin{tabular}{ccc}
\toprule
Symbol & Value & Description\\
\hline
$\alpha$ & 0.2 dB/km & attenuation coefficient\\
$\beta$ & 0.95 & reverse reconciliation efficiency\\
$\varepsilon_{ch}$ & 0.002 & channel noise of phase reference\\
$q$ & 10 & quantization bit number of ADC\\
$R_e$ & 60 dB & finite extinction ratio of AM\\
$d_{dB}$ & 40 & finite dynamic of AM\\
$V_{U_1}$ & 4 & \multirow{2}*{signal modulation variances}\\ 
$V_{U_2}$ & 4 &\\
$\alpha_{Smax_1}$ & $\sqrt{10V_{U_1}}$ & \multirow{2}*{signal maximum amplitudes}\\ 
$\alpha_{Smax_2}$ & $\sqrt{10V_{U_2}}$ &\\ 
\toprule
\end{tabular}
\end{table}
\begin{figure}[!htbp]
\centering
\includegraphics[width=8.5cm]{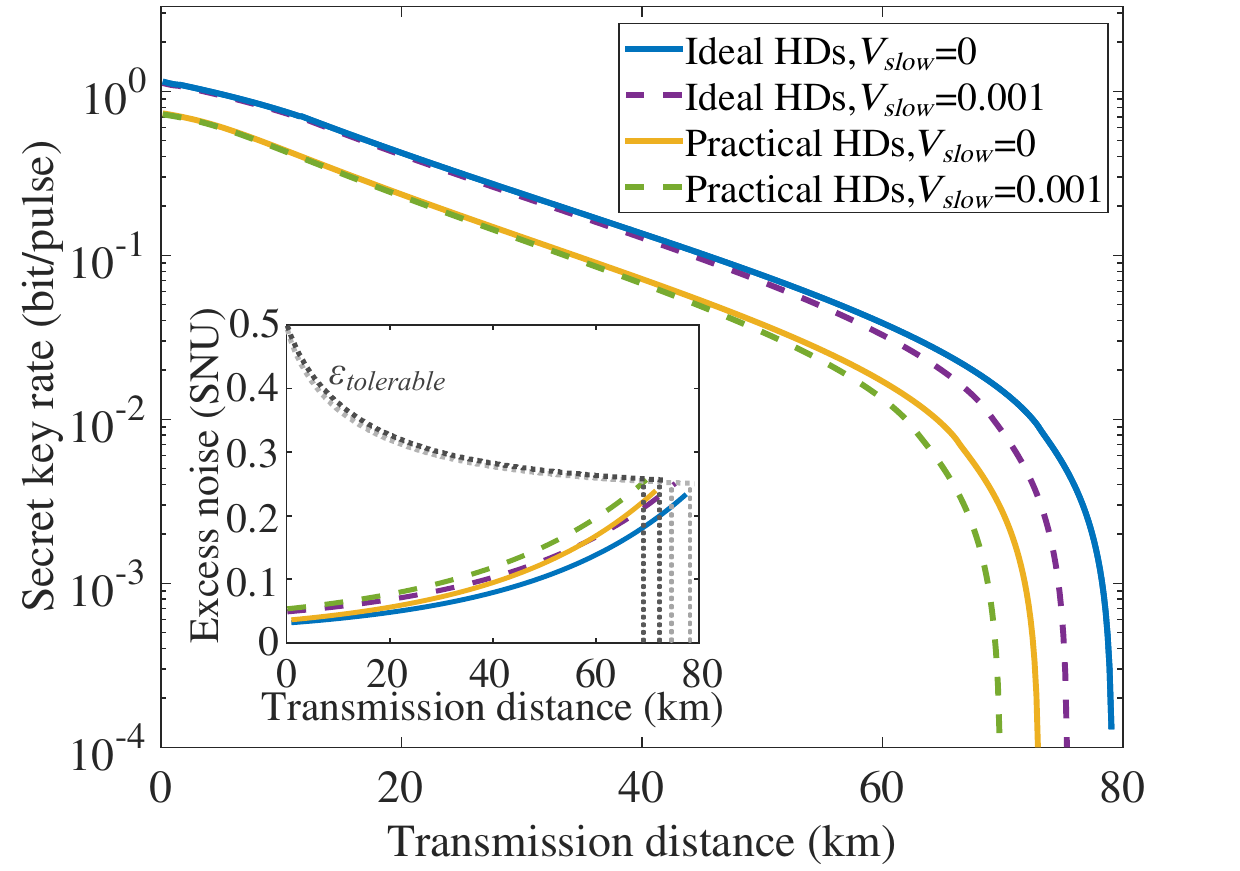}
\caption{Secret key rate of LLO-CVQSS as a function of transmission distance. From right to left, the blue solid line, purple dashed line, yellow solid line and green dashed line respectively show secret key rates with four different situations. Inset depicts the system's total excess noise $\varepsilon$ and tolerable excess noise $\varepsilon_{tolerable}$ as functions of transmission distance.}
\label{fig:Ephase}
\end{figure}

Before performing numerical simulations, several global parameters have to be assigned according to the realistic experimental environment \cite{0902Fossier}. Table \ref{TABLE} presents these global parameters in detail. The asymptotic performance of our proposed LLO-CVQSS scheme is shown in Fig. \ref{fig:Ephase}. Blue solid line depicts that the maximum transmission distance of LLO-CVQSS is close to 80 km when the compensation error of phase slow drift $V_{slow}=0$. This result reveals that the best performance of LLO-CVQSS can be achieved by perfectly performing our proposed phase compensation method. It is worthy noting that the phase fast drift caused by measurement error $\varepsilon_{error}$ cannot be thoroughly eliminated even if HDs are ideal ($\eta=1$ and $v_{el}=0$), since the channel loss $T$ and channel noise $\varepsilon_{ch}$ are existed objectively in every quantum communication system \cite{1905YZhang}. Purple dashed line shows the performance of LLO-CVQSS with ideal HDs in the condition of imperfect phase slow drift compensation ($V_{slow}=0.001$). It indicates that the performance of LLO-CVQSS, especially in terms of maximum transmission distance, will be reduced when the cumulative phase differences $\Delta\theta_{D1}^{delay}$ and $\Delta\theta_{D2}^{delay}$ cannot be precisely estimated. Fortunately, the precision of phase slow drift compensation in our scheme can be well improved by simply increasing the size of the disclosed data used for calculating the estimates. We also investigate the performance of LLO-CVQSS in practical situation. Specifically, yellow solid line and green dashed line respectively show the performance of LLO-CVQSS with practical HDs ($\eta=0.6$ and $v_{el}=0.01$) in the condition of different $V_{slow}$. Obviously, their maximum transmission distance is shorter than that of LLO-CVQSS with ideal HDs, this performance degeneration comes from the increment of the phase measurement noise $\varepsilon_{error}$ as the phase fast drift cannot be ideally compensated in practical situation. Inset shows the system's total excess noise $\varepsilon$ and tolerable excess noise $\varepsilon_{tolerable}$ as functions of transmission distance, it is easy to see that the situation with practical HDs and imperfect phase slow drift compensation will introduce more excess noise to LLO-CVQSS system than other three situations, and the maximum transmission distance for each situation are coincidentally achieved when  $\varepsilon$ equals to $\varepsilon_{tolerable}$. This result matches the result shown in the main figure, and it verifies that the performance of LLO-CVQSS can be well enhanced by the proposed phase compensation method.
%Case of ideal HDs and $V_{slow}=0$ is used in the following analysis.

\begin{figure}[htb]
\centering
\includegraphics[width=8.5cm]{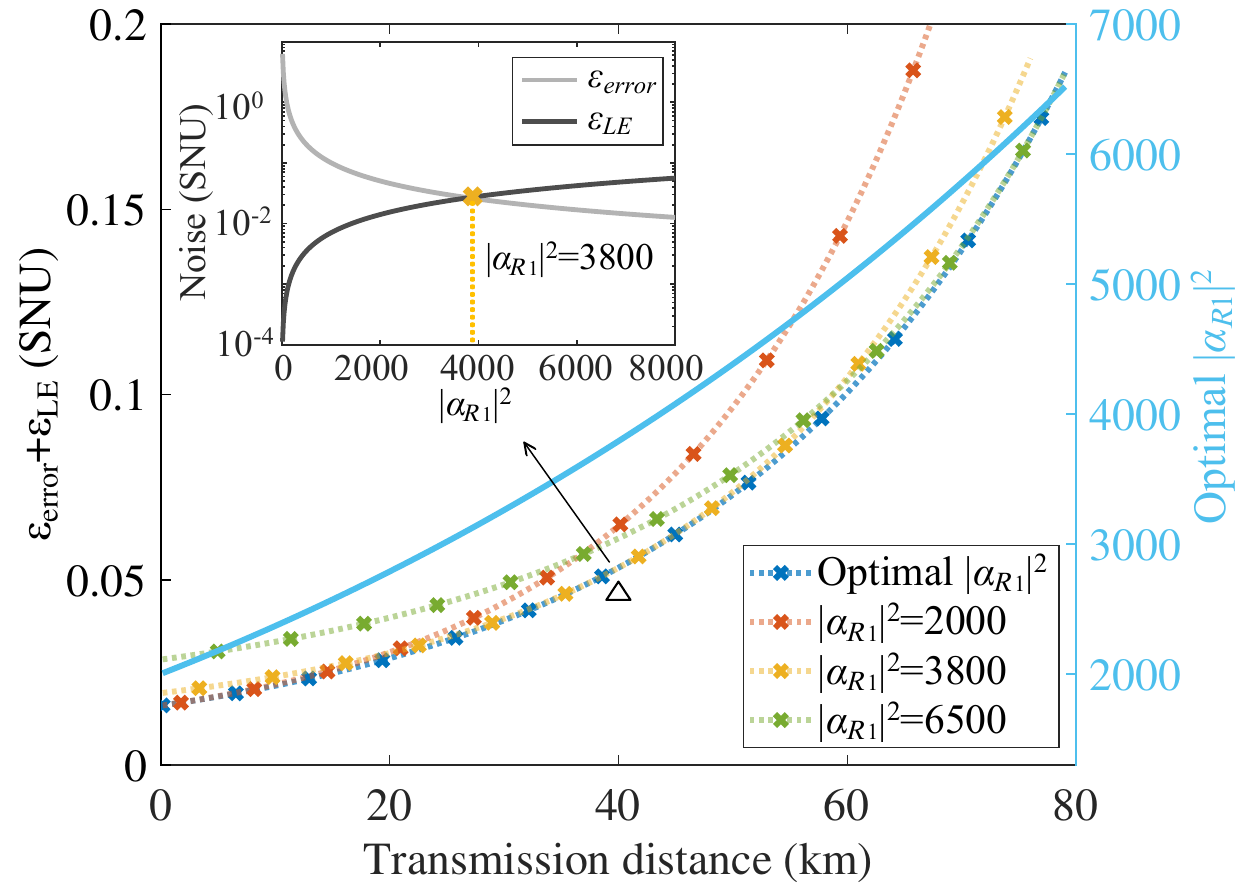}
\caption{The sum of phase measurement noise and photon-leakage noise $\varepsilon_{error}+\varepsilon_{LE}$ as a function of transmission distance in the situation of ideal HDs and $V_{slow}=0$. Blue, orange, yellow and green dotted lines with $\times$ respectively denote user 1's different reference signal amplitudes $|\alpha_{R_1}|^2$. The light blue line represents the optimal $|\alpha_{R_1}|^2$ for each transmission distance. Inset shows $\varepsilon_{error}$ and $\varepsilon_{LE}$ as functions of $|\alpha_{R_1}|^2$ in the condition of setting transmission distance to 40 km.}
\label{fig:aR12}
\end{figure}
Note that the above results is given by optimizing the amplitude of user 1's reference signal $|\alpha_{R_1}|^2$, which is an important setting of LLO-based quantum communication system \cite{1510Soh}. According to the noise model detailed in Sec. \ref{IIC}, the setting of $|\alpha_{R_1}|^2$ is strongly related to photon-leakage noise $\varepsilon_{LE}$ and phase measurement noise $\varepsilon_{error}$. To figure out how $|\alpha_{R_1}|^2$ affects the performance of LLO-CVQSS system, we further plot Fig. \ref{fig:aR12} that demonstrates the sum of $\varepsilon_{error}$ and $\varepsilon_{LE}$ as a function of transmission distance in the situation of ideal HDs and $V_{slow}=0$. It can be easily found that the smallest $\varepsilon_{error}+\varepsilon_{LE}$ (blue dotted line with $\times$) can be obtained by continuously optimizing $|\alpha_{R_1}|^2$ (light blue line) at each transmission distance, rather than setting $|\alpha_{R_1}|^2$ to several fixed values, such as 2000 (red dotted line with $\times$), 3800 (orange dotted line with $\times$), and 6500 (green dotted line with $\times$). Obviously, the best performance of LLO-CVQSS can be achieved with the smallest $\varepsilon_{error}+\varepsilon_{LE}$ since other elements of the total excess noise $\varepsilon$, such as $\varepsilon_{AM}$, $\varepsilon_{ADC}$ and $\varepsilon_{rest}$, will not be affected by $|\alpha_{R_1}|^2$. With Eq.(\ref{e19}) and Eq.(\ref{e22}), we can easily infer that the optimal $|\alpha_{R_1}|^2$ can be obtained when $\varepsilon_{error}$ equals to $\varepsilon_{LE}$ for each distance, this is depicted in the inset which shows an example of how to determine the optimal $|\alpha_{R_1}|^2$ for LLO-CVQSS system with transmission distance is 40 km.

\begin{figure}[htb]
\centering
\includegraphics[width=8.5cm]{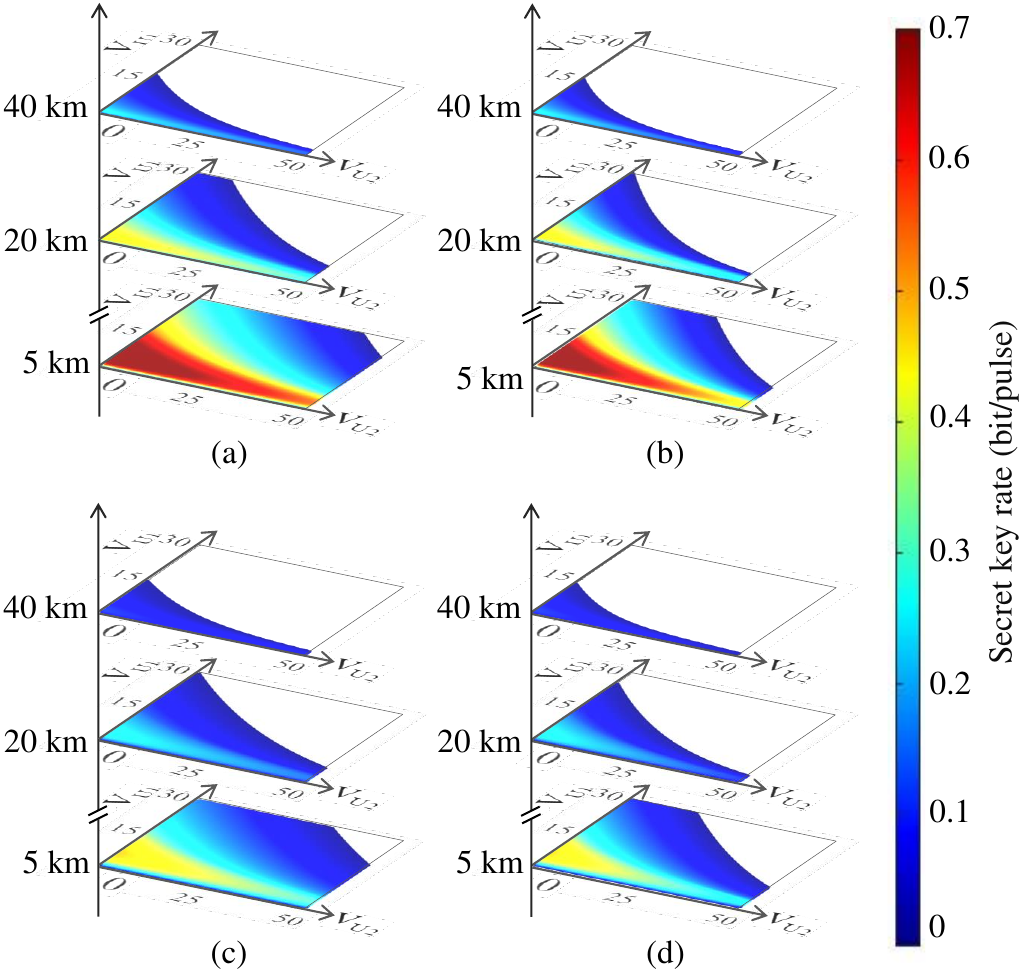}
\caption{Secret key rate of LLO-CVQSS as a function of user 1’s modulation variance $V_{U_1}$ and user 2’s modulation variance $V_{U_2}$ in the condition of (a) ideal HDs and $V_{slow}=0$, (b) ideal HDs and $V_{slow}=0.001$, (c) practical HDs and $V_{slow}=0$ and (d) practical HDs and $V_{slow}=0.001$.}
\label{fig:VA12}
\end{figure}
Another crucial parameter of LLO-CVQSS is modulation variance of quantum signal. Different from the conventional two-party CVQKD that signal modulation is only operated by one user (i.e., Alice) \cite{1802TWang}, all users in LLO-CVQSS have to modulate their respective signal so that GMCSs can be locally prepared. Therefore, how to set a proper modulation variance for each user is important as well. Fig. \ref{fig:VA12} investigates the performance of LLO-CVQSS with different user 1 and user 2's modulation variances at 5 km, 20 km and 40 km. Although the results of the four subplots are different, their trends are similar. In particular, the best secret key rate for LLO-CVQSS at 5 km can be achieved when $V_{U_1}$ and $V_{U_2}$ are respectively limited to the values from approximately 0 to 15 and from approximately 0 to 25, and these ranges will be further narrowed down as transmission distance increases. It indicates that we should choose relatively small value of the modulation variances for optimizing the performance of LLO-CVQSS system. Moreover, the results show that the performance of LLO-CVQSS system is more sensitive to the variation of $V_{U_1}$ than that of $V_{U_2}$, this phenomenon is attributed to the fact that the secret key rate of LLO-CVQSS is theoretically be the secret key rate of the CVQKD link between the farthest user (i.e., user 1) and the dealer, as we should choose the minimal secret key rate among all CVQKD links in \emph{Step 7}. Nevertheless, we point out that the minimal secret key rate of CVQKD links in a practical LLO-CVQSS system must be estimated from realistic data, so it may not belong to the longest CVQKD link.

\begin{figure}[!htbp]
\centering
\includegraphics[width=8.5cm]{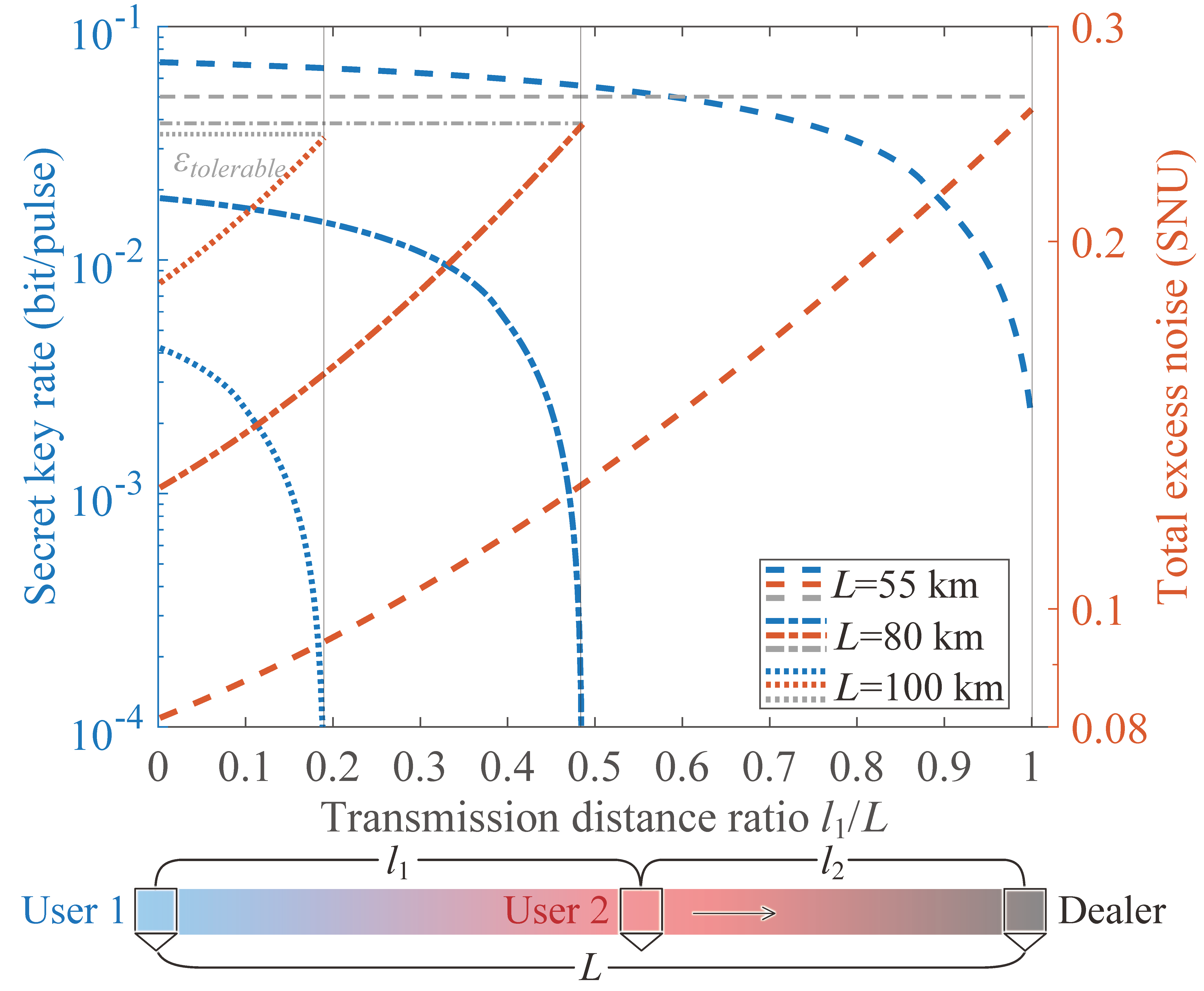}
\caption{Secret key rate (blue lines) and total excess noise $\varepsilon$ (red lines) of LLO-CVQSS as functions of transmission distance ratio $l_1/L$ in the condition of ideal HDs and $V_{slow}=0$. Dashed lines, dotted-dashed lines and dotted lines respectively indicate transmission distances $L$ of 55 km, 80 km and 100 km. Gray lines indicate the tolerable excess noise $\varepsilon_{tolerable}$ for each $L$.}
\label{fig:user2}
\end{figure}
Above analyses are all focused on a symmetric situation that the distance ($l_1$) from user 1 to user 2 and the distance ($l_2$) from user 2 to the dealer are equal, namely user 2 is always located at the midpoint between user 1 and the dealer. Actually, the position of user 2 should be able to flexibly adjust according to his realistic requirement. To explore the impact of this asymmetric situation on the performance of LLO-CVQSS, we plot Fig. \ref{fig:user2} which shows the secret key rate and the total excess noise as functions of transmission distance ratio $l_1/L$ in the situation of ideal HDs and $V_{slow}=0$. On the whole, the secret key rates (blue lines) decrease as transmission distance ratio increases, which reveals that the performance of LLO-CVQSS will be deteriorated as the position of user 2 gradually moves from user 1 to the dealer. This is because total excess noise of LLO-CVQSS system (red lines) raises 
%to the tolerable excess noise $\varepsilon_{tolerable}$ of each $L$
as transmission distance ratio increases. Indeed, this phenomenon can be explained by the noise model of LLO-CVQSS presented in Sec. \ref{IIC}, as reducing $l_2$ will increase user 2's transmittance $T_2$, leading to the increase of each kind of excess noise.
Obviously, the best performance of LLO-CVQSS in each transmission distance can be achieved by moving user 2 as close as possible to user 1. This finding suggests that users are not allowed to be too far apart if one wants to get better performance of LLO-CVQSS system.

\begin{figure}[!htbp]
\centering
\includegraphics[width=8.5cm]{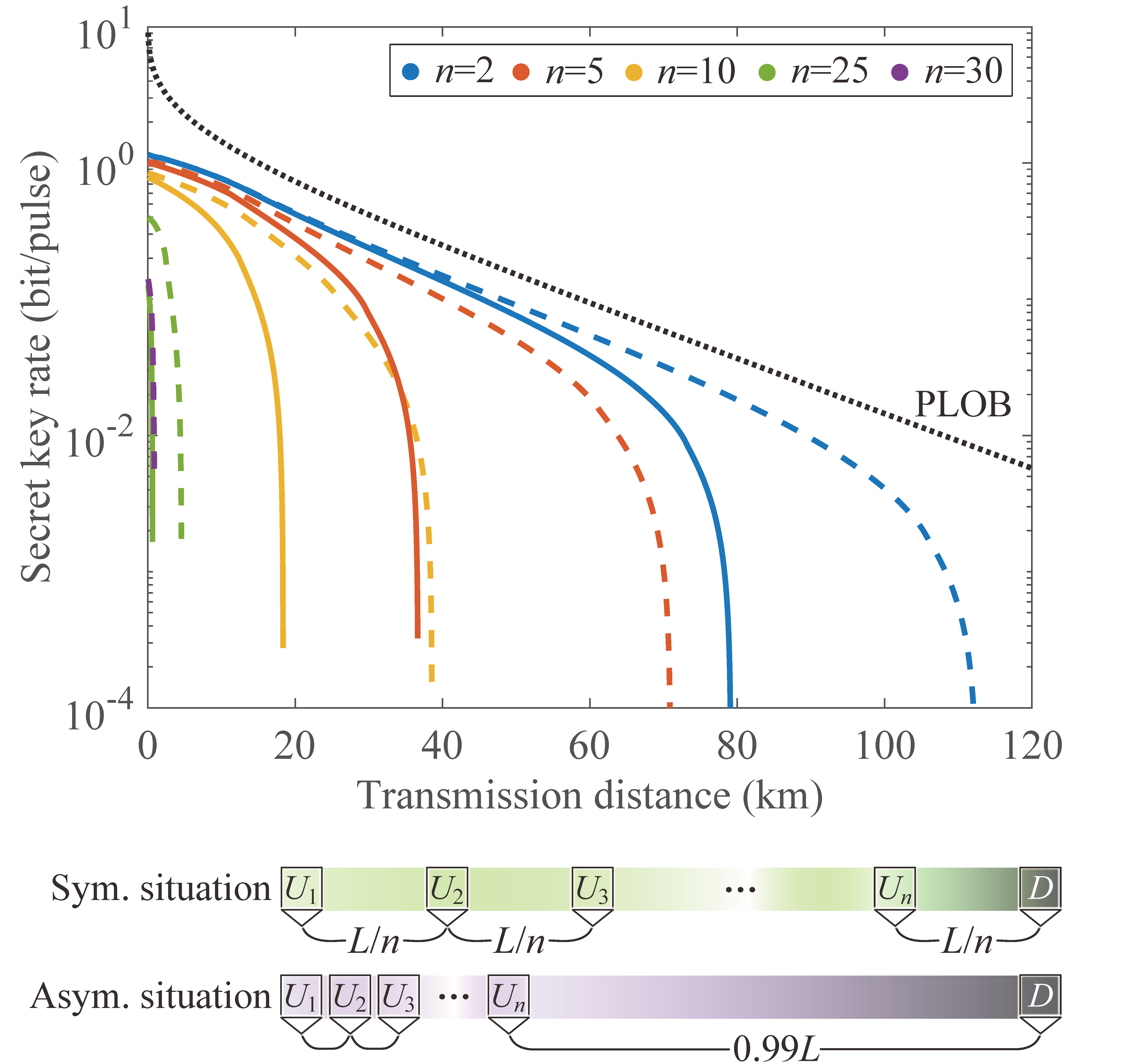}
\caption{Secret key rate of ($n$, $n$) threshold LLO-CVQSS scheme as a function of transmission distance in the condition of ideal HDs and $V_{slow}=0$, modulation variance for each user is set to 4. Solid lines represent symmetric situation and dashed lines represent asymmetric situation. From right to left, solid (dashed) lines indicate the number of users is 2, 5, 10, 25 and 30, respectively. Black dotted line denotes the Piradola-Laurenza-Ottaviani-Banchi (PLOB) bound \cite{1704PLOB}.}
\label{fig:usern}
\end{figure}
Finally, let us consider the user capacity of LLO-CVQSS. Although we only detail the design of (2, 2) threshold LLO-CVQSS scheme in Sec. \ref{II}, ($n$, $n$) threshold LLO-CVQSS scheme can also be implemented by generalizing our proposed ideas. Therefore, it is necessary to investigate the impact of the number of users on LLO-CVQSS system, as shown in Fig.\ref{fig:usern} where both symmetric situation, which all users are located with equal distance intervals, and asymmetric situation, which other users are sequentially located close to user 1, are considered. It can be easily found that the performance of LLO-CVQSS in both situations is reduced as the number of users increases, this is because adding more users will introduce more excess noise to LLO-CVQSS system since $n-1$ users have to be deemed dishonest for security reasons. Moreover, it is also observed that LLO-CVQSS with asymmetric situation (dashed lines) outperforms LLO-CVQSS with symmetric situation (solid lines) especially in terms of maximal transmission distance, revealing that the former is able to support more users than the latter. For example, the maximal number of users of LLO-CVQSS with asymmetric situation reaches 30 while that of LLO-CVQSS with symmetric situation is 25. This result suggests that ($n$, $n$) threshold LLO-CVQSS with asymmetric situation is more beneficial for building large-scale multi-party quantum communication networks.

\section{\label{IV}conclusion\protect}
In this work, we have proposed a practical CVQSS scheme based on LLO (LLO-CVQSS). The proposed LLO-CVQSS scheme waives the necessity that all LOs have to be transmitted through an untrusted channel, so that all LO-aimed attacks can be well defended, greatly improving the practical security of CVQSS system. We first presented the detailed design of LLO-CVQSS in which LO is locally generated by the legitimate party dealer, and then developed a specially designed phase compensation method to eliminate the negative effects caused by fast and slow phase drift. We subsequently constructed a noise model for the proposed LLO-CVQSS and derived its security bound against both eavesdroppers and dishonest users. We detailedly analyzed the asymptotic performance of the proposed LLO-CVQSS and the results showed that LLO-CVQSS with asymmetric situation is able to support 30 users at the same time and its maximal transmission distance reaches 112 km.

In summary, the proposed LLO-CVQSS is not only has the ability to defend itself against all LO-aimed attacks but also has the potential for building large-scale practical quantum communication networks.

\begin{acknowledgments}
This work was supported by the National Natural Science Foundation of China (Grant No. 62101180), Hunan Provincial Natural Science Foundation of China (Grant No. 2022JJ30163).
\end{acknowledgments}

\appendix
\section{\label{Appendix A}Calculation of the correlation}
As analyzed in the main text, to estimate the cumulative phase differences $\Delta\theta_{D1}^{delay}$ and $\Delta\theta_{D2}^{delay}$, user 1 and user 2 need to calculate the correlation with dealer's disclosed data. The correlation between user 1's $x$ quadrature and $x_D'$ is given by
\begin{widetext}\begin{equation}\label{ea1}
\begin{aligned}
\langle x_1x_D'\rangle=&\sqrt{\frac{\eta T_1}2}
[\begin{array}{cc}
    \cos(\Delta\theta_{D1}^{delay}) & \sin(\Delta\theta_{D1}^{delay})
\end{array}]
\Big[\begin{array}{cc}
    \langle x_1x_1\rangle\\
    \langle x_1p_1\rangle
\end{array}\Big]+\\
&\sqrt{\frac{\eta T_2}2}
[\begin{array}{cc}
    \cos(-\theta_2^R+\Delta\theta_{D2}^{delay}) & \sin(-\theta_2^R+\Delta\theta_{D2}^{delay}) 
\end{array}]
\Big[\begin{array}{cc}
    \langle x_1x_2\rangle\\
    \langle x_1p_2\rangle
\end{array}\Big]+\langle x_1x_N'\rangle\\
=&\sqrt{\frac{\eta T_1}2}\cos(\Delta\theta_{D1}^{delay})\langle x_1x_1\rangle+\sqrt{\frac{\eta T_1}2}\sin(\Delta\theta_{D1}^{delay})\langle x_1p_1\rangle+\\
&\sqrt{\frac{\eta T_2}2}\cos(-\theta_2^R+\Delta\theta_{D2}^{delay})\langle x_1x_2\rangle+\sqrt{\frac{\eta T_2}2}\sin(-\theta_2^R+\Delta\theta_{D2}^{delay})\langle x_1p_2\rangle+\langle x_1x_N'\rangle.
\end{aligned}
\end{equation}
\end{widetext}
In the correlations included above, since $x_1$, $p_1$, $x_2$, $p_2$ and $x_N'$ are independent of each other, $\langle x_1p_1\rangle=\langle x_1x_2\rangle=\langle x_1p_2\rangle=\langle x_1x_N'\rangle=0$. Hence Eq.(\ref{ea1}) can be revised to
\begin{equation}\label{ea2}
\langle x_1x_D'\rangle=\sqrt{\frac{\eta T_1}2}\cos(\Delta\theta_{D1}^{delay})\langle x_1x_1\rangle.
\end{equation}
The correlation between user 1's $p$ quadrature and $x_D'$ is similarly given by
\begin{widetext}\begin{equation}\label{ea3}
\begin{aligned}
\langle p_1x_D'\rangle=&\sqrt{\frac{\eta T_1}2}
[\begin{array}{cc}
    \cos(\Delta\theta_{D1}^{delay}) & \sin(\Delta\theta_{D1}^{delay})
\end{array}]
\Big[\begin{array}{cc}
    \langle p_1x_1\rangle\\
    \langle p_1p_1\rangle
\end{array}\Big]+\\
&\sqrt{\frac{\eta T_2}2}
[\begin{array}{cc}
    \cos(-\theta_2^R+\Delta\theta_{D2}^{delay}) & \sin(-\theta_2^R+\Delta\theta_{D2}^{delay}) 
\end{array}]
\Big[\begin{array}{cc}
    \langle p_1x_2\rangle\\
    \langle p_1p_2\rangle
\end{array}\Big]+\langle p_1x_N'\rangle\\
=&\sqrt{\frac{\eta T_1}2}\cos(\Delta\theta_{D1}^{delay})\langle p_1x_1\rangle+\sqrt{\frac{\eta T_1}2}\sin(\Delta\theta_{D1}^{delay})\langle p_1p_1\rangle+\\
&\sqrt{\frac{\eta T_2}2}\cos(-\theta_2^R+\Delta\theta_{D2}^{delay})\langle p_1x_2\rangle+\sqrt{\frac{\eta T_2}2}\sin(-\theta_2^R+\Delta\theta_{D2}^{delay})\langle p_1p_2\rangle+\langle p_1x_N'\rangle\\
=&\sqrt{\frac{\eta T_1}2}\sin(\Delta\theta_{D1}^{delay})\langle p_1p_1\rangle,
\end{aligned}
\end{equation}
\end{widetext}
where $\langle p_1p_1\rangle=\langle x_1x_1\rangle$. With Eq.(\ref{ea2}) and Eq.(\ref{ea3}), user 1 can have the estimated value $\Delta\hat{\theta}_{D1}^{delay}$ as Eq.(\ref{e14}) in the main text, which is prior to parameter estimation (without knowing its channel transmittance $T_1$).

For user 2, the correlation is performed with its data $x_2'$ after the first phase rotation, calculated by
\begin{widetext}\begin{equation}\label{ea4}
\begin{aligned}
\langle x_2'x_D'\rangle=&\sqrt{\frac{\eta T_1}2}
[\begin{array}{cc}
    \cos(-\theta_2^R)\cos(\Delta\theta_{D1}^{delay}) & 
    \cos(-\theta_2^R)\sin(\Delta\theta_{D1}^{delay})
\end{array}]
\Big[\begin{array}{cc}
    \langle x_2x_1\rangle\\
    \langle x_2p_1\rangle
\end{array}\Big]+\\
&\sqrt{\frac{\eta T_2}2}
[\begin{array}{cc}
    \cos(-\theta_2^R)\cos(-\theta_2^R+\Delta\theta_{D2}^{delay}) &
    \cos(-\theta_2^R)\sin(-\theta_2^R+\Delta\theta_{D2}^{delay}) 
\end{array}]
\Big[\begin{array}{cc}
    \langle x_2x_2\rangle\\
    \langle x_2p_2\rangle
\end{array}\Big]+\cos(-\theta_2^R)\langle x_2x_N'\rangle+\\
&\sqrt{\frac{\eta T_1}2}
[\begin{array}{cc}
    \sin(-\theta_2^R)\cos(\Delta\theta_{D1}^{delay}) & 
    \sin(-\theta_2^R)\sin(\Delta\theta_{D1}^{delay})
\end{array}]
\Big[\begin{array}{cc}
    \langle p_2x_1\rangle\\
    \langle p_2p_1\rangle
\end{array}\Big]+\\
&\sqrt{\frac{\eta T_2}2}
[\begin{array}{cc}
    \sin(-\theta_2^R)\cos(-\theta_2^R+\Delta\theta_{D2}^{delay}) &
    \sin(-\theta_2^R)\sin(-\theta_2^R+\Delta\theta_{D2}^{delay}) 
\end{array}]
\Big[\begin{array}{cc}
    \langle p_2x_2\rangle\\
    \langle p_2p_2\rangle
\end{array}\Big]+\sin(-\theta_2^R)\langle p_2x_N'\rangle\\
=&\sqrt{\frac{\eta T_1}2}\cos(-\theta_2^R)\cos(\Delta\theta_{D1}^{delay})\langle x_2x_1\rangle+\sqrt{\frac{\eta T_1}2}\cos(-\theta_2^R)\sin(\Delta\theta_{D1}^{delay})\langle x_2p_1\rangle+\\
&\sqrt{\frac{\eta T_2}2}\cos(-\theta_2^R)\cos(-\theta_2^R+\Delta\theta_{D2}^{delay})\langle x_2x_2\rangle+\sqrt{\frac{\eta T_2}2}\cos(-\theta_2^R)\sin(-\theta_2^R+\Delta\theta_{D2}^{delay})\langle x_2p_2\rangle+\cos(-\theta_2^R)\langle x_2x_N'\rangle+\\
&\sqrt{\frac{\eta T_1}2}\sin(-\theta_2^R)\cos(\Delta\theta_{D1}^{delay})\langle p_2x_1\rangle+\sqrt{\frac{\eta T_1}2}\sin(-\theta_2^R)\sin(\Delta\theta_{D1}^{delay})\langle p_2p_1\rangle+\\
&\sqrt{\frac{\eta T_2}2}\sin(-\theta_2^R)\cos(-\theta_2^R+\Delta\theta_{D2}^{delay})\langle p_2x_2\rangle+\sqrt{\frac{\eta T_2}2}\sin(-\theta_2^R)\sin(-\theta_2^R+\Delta\theta_{D2}^{delay})\langle p_2p_2\rangle+\sin(-\theta_2^R)\langle p_2x_N'\rangle\\
=&\sqrt{\frac{\eta T_2}2}\cos(-\theta_2^R)\cos(-\theta_2^R+\Delta\theta_{D2}^{delay})\langle x_2x_2\rangle+\sqrt{\frac{\eta T_2}2}\sin(-\theta_2^R)\sin(-\theta_2^R+\Delta\theta_{D2}^{delay})\langle p_2p_2\rangle\\
=&\sqrt{\frac{\eta T_2}2}\cos(-\theta_2^R+\theta_2^R-\Delta\theta_{D2}^{delay})\langle x_2x_2\rangle\\
=&\sqrt{\frac{\eta T_2}2}\cos(\Delta\theta_{D2}^{delay})\langle x_2x_2\rangle,
\end{aligned}
\end{equation}
\end{widetext}
where $\langle x_2x_2\rangle=\langle p_2p_2\rangle$. And the correlation between $p_2'$ and $x_D'$ is calculated by
\begin{widetext}\begin{equation}\label{ea5}
\begin{aligned}
\langle p_2'x_D'\rangle=&\sqrt{\frac{\eta T_1}2}
[\begin{array}{cc}
    -\sin(-\theta_2^R)\cos(\Delta\theta_{D1}^{delay}) & 
    -\sin(-\theta_2^R)\sin(\Delta\theta_{D1}^{delay})
\end{array}]
\Big[\begin{array}{cc}
    \langle x_2x_1\rangle\\
    \langle x_2p_1\rangle
\end{array}\Big]+\\
&\sqrt{\frac{\eta T_2}2}
[\begin{array}{cc}
    -\sin(-\theta_2^R)\cos(-\theta_2^R+\Delta\theta_{D2}^{delay}) &
    -\sin(-\theta_2^R)\sin(-\theta_2^R+\Delta\theta_{D2}^{delay}) 
\end{array}]
\Big[\begin{array}{cc}
    \langle x_2x_2\rangle\\
    \langle x_2p_2\rangle
\end{array}\Big]-\sin(-\theta_2^R)\langle x_2x_N'\rangle+\\
&\sqrt{\frac{\eta T_1}2}
[\begin{array}{cc}
    \cos(-\theta_2^R)\cos(\Delta\theta_{D1}^{delay}) & 
    \cos(-\theta_2^R)\sin(\Delta\theta_{D1}^{delay})
\end{array}]
\Big[\begin{array}{cc}
    \langle p_2x_1\rangle\\
    \langle p_2p_1\rangle
\end{array}\Big]+\\
&\sqrt{\frac{\eta T_2}2}
[\begin{array}{cc}
    \cos(-\theta_2^R)\cos(-\theta_2^R+\Delta\theta_{D2}^{delay}) &
    \cos(-\theta_2^R)\sin(-\theta_2^R+\Delta\theta_{D2}^{delay}) 
\end{array}]
\Big[\begin{array}{cc}
    \langle p_2x_2\rangle\\
    \langle p_2p_2\rangle
\end{array}\Big]+\cos(-\theta_2^R)\langle p_2x_N'\rangle\\
=&-\sqrt{\frac{\eta T_1}2}\sin(-\theta_2^R)\cos(\Delta\theta_{D1}^{delay})\langle x_2x_1\rangle-\sqrt{\frac{\eta T_1}2}\sin(-\theta_2^R)\sin(\Delta\theta_{D1}^{delay})\langle x_2p_1\rangle\\
&-\sqrt{\frac{\eta T_2}2}\sin(-\theta_2^R)\cos(-\theta_2^R+\Delta\theta_{D2}^{delay})\langle x_2x_2\rangle-\sqrt{\frac{\eta T_2}2}\sin(-\theta_2^R)\sin(-\theta_2^R+\Delta\theta_{D2}^{delay})\langle x_2p_2\rangle-\sin(-\theta_2^R)\langle x_2x_N'\rangle\\
&+\sqrt{\frac{\eta T_1}2}\cos(-\theta_2^R)\cos(\Delta\theta_{D1}^{delay})\langle p_2x_1\rangle+\sqrt{\frac{\eta T_1}2}\cos(-\theta_2^R)\sin(\Delta\theta_{D1}^{delay})\langle p_2p_1\rangle\\
&+\sqrt{\frac{\eta T_2}2}\cos(-\theta_2^R)\cos(-\theta_2^R+\Delta\theta_{D2}^{delay})\langle p_2x_2\rangle+\sqrt{\frac{\eta T_2}2}\cos(-\theta_2^R)\sin(-\theta_2^R+\Delta\theta_{D2}^{delay})\langle p_2p_2\rangle+\cos(-\theta_2^R)\langle p_2x_N'\rangle\\
=&-\sqrt{\frac{\eta T_2}2}\sin(-\theta_2^R)\cos(-\theta_2^R+\Delta\theta_{D2}^{delay})\langle x_2x_2\rangle+\sqrt{\frac{\eta T_2}2}\cos(-\theta_2^R)\sin(-\theta_2^R+\Delta\theta_{D2}^{delay})\langle p_2p_2\rangle\\
=&-\sqrt{\frac{\eta T_2}2}\sin(-\theta_2^R+\theta_2^R-\Delta\theta_{D2}^{delay})\langle x_2x_2\rangle\\
=&\sqrt{\frac{\eta T_2}2}\sin(\Delta\theta_{D2}^{delay})\langle x_2x_2\rangle.
\end{aligned}
\end{equation}
\end{widetext}
Then user 2 can obtain the estimated value $\Delta\hat{\theta}_{D2}^{delay}$ in Eq. (\ref{e15}) before parameter estimation.
\section{\label{Appendix B}Secret key rate of LLO-CVQSS}
The lower bound of secret key rate for LLO-CVQSS can be represented by
\begin{equation}\label{eb1}
R=\beta I_{UD}-\chi_{DE},
\end{equation}
where $\beta$ is the reverse reconciliation efficiency, $I_{UD}$ is the Shannon mutual information between user 1 and the dealer, and $\chi_{DE}$ is the maximum information available to eavesdropper on dealer's measurement.

Combined with excess noise analyzed in Sec. \ref{IIC}, the channel-added noise referred to the channel input is given by
\begin{equation}\label{eb2}
\chi_{line}=\frac1{T_1}-1+\varepsilon,
\end{equation}
and the noise added by dealer's heterodyne detector is defined as
\begin{equation}\label{eb3}
\chi_{het}=\frac{2-\eta+2v_{el}}\eta,
\end{equation}
where $\eta$ is the detection efficiency and $v_{el}$ is the electronic noise of imperfect detector. Thus the overall noise referred to the channel input can be denoted as
\begin{equation}\label{eb4}
\chi_{tot}=\chi_{line}+\frac{\chi_{het}}{T_1}.
\end{equation}
For now the Shannon mutual information $I_{UD}$ is calculated by
\begin{equation}\label{eb5}
I_{UD}=\log_2{\frac{V+\chi_{tot}}{1+\chi_{tot}}},
\end{equation}
where $V=V_{U_1}+1$.

The term $\chi_{DE}$ can be simply calculated as \cite{0710Lodewyck}
\begin{equation}\label{eb6}
\chi_{DE}=\sum_{j=1}^2G(\frac{\lambda_j-1}2)-\sum_{j=3}^5G(\frac{\lambda_j-1}2),
\end{equation}
where $G(x)=(x+1)\log_2(x+1)-x\log_2x$, and the symplectic eigenvalues are
\begin{equation}\label{eb7}
\lambda_{1,2}^2=\frac12[A\pm\sqrt{A^2-4B}],
\end{equation}
where
\begin{equation}\label{eb8}
A=V^2(1-2T_1)+2T_1+T_1^2(V+\chi_{line})^2,
\end{equation}
\begin{equation}\label{eb9}
B=T_1^2(V\chi_{line}+1)^2,
\end{equation}
\begin{equation}\label{eb10}
\lambda_{3,4}^2=\frac12[C\pm\sqrt{C^2-4D}],
\end{equation}
\begin{equation}\label{eb11}
\lambda_5=1,
\end{equation}
where
\begin{equation}\label{eb12}
\begin{aligned}
C=&\frac1{[T_1(V+\chi_{tot})]^2}\{A\chi_{het}^2+B+1+\\&2\chi_{het}[V\sqrt B+T_1(V+\chi_{line})]+2T_1(V^2-1)\},
\end{aligned}
\end{equation}
\begin{equation}\label{eb13}
D=\Big[\frac{V+\sqrt B\chi_{het}}{T_1(V+\chi_{tot})}\Big]^2.
\end{equation}
%\nocite{*}
\bibliography{reference}
\end{document}